\documentclass[11pt,twocolumn, noshowpacs,nofootinbib,notitlepage,amsmath]{revtex4-1}
\allowdisplaybreaks
\usepackage{graphicx,color}
\usepackage[colorlinks=true,citecolor=blue,linkcolor=blue,urlcolor=blue]{hyperref}
\usepackage[charter]{mathdesign}
\DeclareSymbolFontAlphabet{\mathcal}{symbols}
\DeclareSymbolFont{symbols}{OMS}{xmdcmsy}{m}{n}
\DeclareSymbolFont{largesymbols}{OMX}{xmdcmex}{m}{n}
\SetSymbolFont{symbols}{bold}{OMS}{xmdcmsy}{b}{n}

\begin{document}  
\title{\color{blue}\Large Dynamical Generation of the Weak Scale and Inflation in High-Scale Supersymmetry}
\author{Sibo Zheng}
\email{sibozheng.zju@gmail.com}
\affiliation{Department of Physics, Chongqing University, Chongqing 401331, P. R. China}
\begin{abstract}
Combination of Plank and BICEP2 data reported that
 tensor to scalar ratio $r\simeq 0.16$ and scalar spectral index $n_{s}\simeq 0.96$. 
In this short note, it is shown that chaotic inflation with quadratic potential,
which perfectly accounts for present fit to $r$ and $n_s$, 
and relatively heavy Higgs mass for fine-tuned MSSM,
are both suggested to be consequences of high-scale supersymmetry (SUSY) breaking.
Following this idea, we find that inflaton mass is of same order as soft mass, 
flatness of inflation potential doesn't need a shift symmetry,
and scale of inflation energy density automatically agrees with SUSY-breaking scale.
A simple realization of such high-scale SUSY breaking, 
in which inflaton with tri-linear superpotential serves as {\it modulus},
is discussed in terms of ISS-like model. 
\end{abstract}
\maketitle

\section{Introduction }

To date cosmic microwave background (CMB) is few of useful windows for our study of early universe. 
For the fit to CMB temperature anisotropy and polarization,
six-parameter $\Lambda$CDM is broadly used as base model.
This set of cosmological parameters includes baryon density $\Omega_{B}$, cold dark matter density $\Omega_{c}$, 
the curvature fluctuation amplitude $A_s$, scalar spectral index $n_s$, and reionization optical depth $\tau$.
Plank excludes the scale invariant case robustly and confirms deviation from  $n_{s}=1$ to red value $n_{s}\simeq 0.96$ \cite{1303.5076,1303.5082}.
Plank also confirmed that only one parameter fit with $n_s$ is excluded. 
This result encouraged Plank further for two-parameter fit (with tensor-scalar ratio $r$ included),
and it gives rise a upper bound $r<0.11$ at 95 \% CL \cite{1303.5076,1303.5082} by using the WMAP data together.
Recently, BICEP2 collaboration reports a direct detection of gravitational wave for the first time.
The fit to BICEP2 data leads to $r\simeq 0.2^{+0.07}_{-0.05}$ at 95 \% CL \cite{1403.3985}.
After a subtraction of the best-available foreground model 
BICEP2 data suggests that \footnote{
The fits to $r$ from Plank and BICEP2 collaboration get {\it moderate} tension with each other, 
roughly of 3$\sigma$ level.
The improved BICEP2 data released later will probably tell us whether the tension is due to systematic uncertainty or new physics is needed to reconcile this problem. 
Even so at present stage a few efforts \cite{recipes} have been devoted to discuss this discrepency.},
\begin{eqnarray}{\label{randns}}
r\simeq 0.16^{+0.06}_{-0.05},~~n_{s}-1\simeq 0.960.
\end{eqnarray}
at 68\% CL.

Parameters such as spectrum index $n_s$ (scalar), $n_t$ (tensor),
their runnings $dn_{s(t)}/d\ln k$ 
and the running of these runnings can be explicitly determined,
if one adopts the small inflation condition.
For single field inflation we have \cite{1303.5082},
\begin{eqnarray}{\label{theoreticfit1}}
n_{s}-1\simeq 2\eta_{V}-6\epsilon_{V},~~~
n_{t}\simeq -2\epsilon_{V}.
\end{eqnarray}
with $r$ is given by,
\begin{eqnarray}{\label{r}}
r\simeq16\epsilon_{V}\simeq -8n_{t}
\end{eqnarray}
where 
\begin{eqnarray}{\label{slowrollpa}}
\epsilon_{V}&\equiv& M^{2}_{P}V^{2}_{,\phi}/2V^{2}, \nonumber\\
\eta_{V}&\equiv& M^{2}_{P}V_{,\phi\phi}/V.
\end{eqnarray}
Here $V_{,\phi}$ and $V_{,\phi\phi}$  refers to the first and second derivative of potential over inflaton $\phi$ respectively, $M_{P}$ is the reduced Plank scale. 
Using these quantitative relations Eq.(\ref{theoreticfit1})-(\ref{r}),
inflation models (For a review on inflation models, see, .e.g. \cite{models}.)
can be directly examined in light of measurements on $r$ and $n_{s}$ by Plank and BICEP2 data,
as analyzed in original paper of Plank \cite{1303.5082}.

The combination of Plank and BICEP2 leads to two important implications for slow roll single field inflation, independent of explicit inflaton potential analyzed.
First, the excursion of inflation is directly related to the measurement on $r$ as,
\begin{eqnarray}{\label{Lythbound}}
\frac{\Delta\phi}{M_{P}}\geq \mathcal{O}(1)\times \sqrt{\frac{r}{0.01}}
\end{eqnarray}
for e-foldings $N\sim 50-60$,
which is known as Lyth bound \cite{Lyth} in slow roll inflation.
$r\simeq 0.16$ suggests that inflation starts from super-Plankian region $\phi>M_{P}$.
Second, $r$ is also directly related to the energy scale of inflation through,
\begin{eqnarray}{\label{scale}}
V_{0}^{1/4}\simeq 2\times10^{16} \left(\frac{r}{0.20}\right)^{1/4}{\it GeV}.
\end{eqnarray}
which suggests that this scale is near the Grand unification (GUT) scale.
This is  a solid evidence for the existence of new physical scale between weak and Plank scale.

A common view from high energy particle physics is 
SUSY breaking is probably broken at high scale $> 10^{10}$ GeV \cite{highscalesusy},
due to the absence of scalar squark signal and relatively large Higgs mass for the minimal supersymmetric model.
If so, it is probably not a coincidence that both BICEP2 and LHC experiments point to large new physical scales nearly of same order.
In this paper, we will take the assumption that {\it inflaton mass is of same order as scalar soft mass scale},
and study high-scale SUSY breaking as the same origin of inflation and Higgs mass.

The paper is organized as follows.
In section II, we consider chaotic inflation, 
in which quadratic potential for inflaton is found to be in perfect agreement with BICEP2 data,
with inflaton mass $m_{\phi}\simeq 10^{13}$ GeV.
In section III, we discuss aspects of model building based on high-scale SUSY breaking.
We use ISS-like model as an illustration.
Inflaton is embedded into this SUSY breaking model as a {\it modulus}.
If so, the scale of inflation energy density automatically agrees with supersymmetry breaking scale,
and the inflation is under theoretic control in terms of electric-magnetic duality.
We find that the SUSY phenomenological features, 
with arrangement of dynamical scales as hinted by the BICEP2 data,
are totally consistent with the present status of SUSY at the LHC 2013 data.

\section{$\Lambda$CDM + $r$ as Base Model}
In this section we discuss the fit for $\Lambda$CDM + $r$ model to both Plank and BICEP2 data.
Take the central values measured for illustration.
Substituting $r\simeq 0.16$, $n_{s}=0.96$ into Eq.(\ref{theoreticfit1}) - (\ref{r}) leads to
\begin{eqnarray}{\label{fits1}}
\epsilon_{V} \simeq 0.01,~~
\eta_{V} \simeq 0.01.
\end{eqnarray}
and $n_{t}\simeq-0.02$ for consistency.

Three important observations follow from Eq.(\ref{fits1}).
At first, the negative tensor index $n_t$ is a smoking gun for such fit,
however it is probably too small to detect.
Second, for chaotic inflation with potential $V\sim \lambda_{n} M^{4}(\phi/M)^{n}$,
where $M$ characterizing the validity of inflation potential is below the Plank scale,
we have 
\begin{eqnarray}{\label{chaotic}}
n_{s}-1\simeq -\left(1+\frac{2}{n}\right)\frac{r}{8}
\end{eqnarray}
from which quadratic potential \cite{Linde1983} gives rise to perfect fit to combination of Plank and BICEP2 data,
as noticed in \cite{1303.5082} and simulated works \cite{quadratic}. 
In summary, the fit based on $\Lambda$CDM +$r$  favors quadratic potential,
\begin{eqnarray}{\label{quadratic}}
V=\frac{1}{2}m_{\phi}^{2}\phi^{2}
\end{eqnarray}
with inflaton mass $m_{\phi}\simeq 10^{13} $ GeV,
which is determined from the equality $m^{2}_{\phi}=\eta_{V}V/M^{2}_{P}$.

Note that for large deviation from $n=2$ it requires that 
Yukawa coupling $\lambda_{n}$ must be sufficiently suppressed 
in order to arrange for a small amplitude of density fluctuation and satisfy the slow roll condition.
For quadratic potential, however,
it automatically ensures that potential energy density is sub-Plankian.
In this sense, common view \cite{1403.7323} on the need of a shift symmetry to provide suppression isn't necessary. 

There is also another common view from viewpoint of particle physics.
The absence of SUSY 
\footnote{Due to the appearance of new physical scale near GUT scale,
the quadratic divergence involving standard model like Higgs scale discovered by the LHC 
must be resolved. SUSY is still few of promising frameworks on board.} 
signals at the Large Hardron collider (LHC) implies that 
SUSY probably breaks at some high $\geq 10^{10}$ GeV other than low scale,
and {\it moderate} fine tunings related to electroweak symmetry breaking have to be taken.
If so, the SUSY-breaking scale is rather close to the inflation energy density scale,
as hinted by the Plank, BICEP2 and LHC experiment.
This is probably not a coincidence.
In the next section, we will discuss quadratic potential Eq.(\ref{quadratic}),
with $m_{\phi}$ equal to scalar soft mass \cite{1403.6081,1403.8138,1405.0068},
which is a consequence of high-scale SUSY breaking as the same origin of inflation and Higgs mass.

\section{ Model Building}

\subsection{Dynamical Scales of SUSY breaking }
As discussed above, 
the inflaton mass $m_{\phi}$ is probably related to the SUSY-breaking scale $\sqrt{F}$ \cite{1403.6081,1403.8138}.
Comparing $V\simeq F^{2}$ with Eq.(\ref{scale}) it follows immediately
\begin{eqnarray}{\label{F}}
\sqrt{F}\sim 10^{16} GeV \sim 10^{3} m_{\phi}.
\end{eqnarray}
Given SUSY breaking effect mediated at scale $M_{*}\sim 3 M_{P}$,  we have from Eq.(\ref{F})
\begin{eqnarray}{\label{scales}}
m_{soft}\sim \frac{F}{M_{*}} \sim m_{\phi}
\end{eqnarray}
The scalar soft mass spectrum Eq.(\ref{scales}) is in agreement with the present status of LHC data,
including relatively large Higgs mass around 126 GeV for non-standard minimal SUSY model \footnote{In the second reference of \cite{highscalesusy}, the fine tuned condition for 
providing $126\pm 3$ GeV is that $\det\mathcal{M}\simeq 0$ at scale $m_{\phi}$, $\mathcal{M}$ being the mass squared matrix for Higgs doublets.
This requirement can be used as a constraint on SUSY-breaking mechanism,
which is  beyond our scope in this note.}.
The obscure to model building is that we lose the theoretic control on analysis of SUSY breaking when $M_{*}\geq M_{P}$.

But ISS-like SUSY breaking model \cite{ISS} is an exception.
We can ensure there allows an electric-magnetic dual for strongly coupled super-Yang-Mills (SYM) gauge theory \footnote{Recently, it is also proposed in \cite{1403.4536} that chaotic inflation potential with power law index $n\leq 1$ is generated through strongly dynamics of SYM .}
at scale $\Lambda$, 
by arranging the number of bi-fundamental matter superfields $Q_{i}$, $N_f$ 
and group quantum number $N_{c}$ in the range $N_{c}+1\leq N_{f}<3N_{c}/2$.
For $\Lambda$ near $\sim M_{P}$,
we can use weak magnetic description to explore this model.
If we introduce a tree-level mass $m_{\phi} Q_{i}\tilde{Q}_{i}$ into ISS model, 
which is reliable for electric-magnetic duality as $m_{\phi}<< \Lambda$,  
then SYM theory  spontaneously breaks SUSY with 
\begin{eqnarray}{\label{F2}}
F\simeq m_{\phi}\Lambda \simeq \left(10^{3} \times m_{\phi}\right)^{2}, ~~\Lambda\simeq 5 M_{P}.
\end{eqnarray}
which {\it naturally} explains quantitative relation between $\sqrt{F}$ and $m_{\phi}$ as shown in Eq.(\ref{F}) if $\Lambda$ is identified as the same order of $\sim M_{*}$.
As a result, the energy density $V^{1/4}\simeq \sqrt{F}\sim 10^{16}$ GeV simultaneously.
This can be consistent with the large field inflation Eq.(\ref{Lythbound}),
once the inflaton is identified as modulus of SUSY breaking vacuum.
This will be addressed in the next section.

The consequences from viewpoint of particle physics phenomenology are as follows.
At first,  the super heavy scalar soft mass spectrum in Eq.(\ref{scales}) predicts the absence of them at the LHC.
Second, the Higgs mass of non-standard minimal SUSY model obtains significant renormalization group effects,
which can explain the Higgs mass $\sim$126 GeV. 
However, some fine tunings such as $\tan\beta\simeq 1$ at the soft scalar mass scale should be imposed so as to produce the correct Higgs mass. 
Finally, gauginos heavier than sfermion are favored by the fit to Higgs mass,
although $R$ symmetry needed for SUSY breaking forbids the generation of gaugino mass.

In summary, SUSY phenomenological features from such ISS-like model,
with arrangement of dynamical scales as hinted by the BICEP2 data,
are totally consistent with the present status of SUSY at the LHC 2013 data.\\
 
\subsection{Inflaton as Modulus of SUSY Breaking Vacuum}
Now we discuss plausible realization for inflaton as a modulus of SUSY breaking vacuum.
The superpotential originated from ISS-like SUSY breaking is specific example of generic O’ Raifeartaigh (O' R) models\footnote{Canonical Kahler potential for both inflaton and other fields are assumed in this paper.
High-dimensional operators induced by quantum gravity in the Kahler potential are assumed small.},
\begin{eqnarray}{\label{superpotential1}}
W=FX+(X+\lambda_{ij}M_{*})\phi_{i}\phi_{j}+\cdots.
\end{eqnarray}
which will be used as starting point in this subsection.
For SUSY breaking models defined as in Eq.(\ref{superpotential1}),
$F$ and $M_{*}$ are identified as  dynamical scales involving SUSY breaking discussed above.
How to embedding inflaton into Eq.(\ref{superpotential1}) as  a modulus is our concern in what follows. 

As well known, for generic O' R model defined in Eq.(\ref{superpotential1}) 
the SUSY breaking vacuum is located at $\phi_{i}=0$ and $X$ arbitrary, 
with $F$-terms given by,
\begin{eqnarray}{\label{Fterm}}
F_{X}=F,~~~~~F_{\phi_{i}}=0.
\end{eqnarray}
As for the pseudo-moduli $X$ , the sign of its
mass squared can be either positive or negative,
which is crucially related to $R$-charge assignments on $\phi_i$ and $X$ \cite{0703196}.
For the case where there are no $R$-charges other than 0 and 2, 
$m^{2}_{X}>0$ and $R$ symmetry is restored at the local minimum of effective potential.
If there are additional $R$ charges other than 0 or 2,
it is possible that $m^{2}_{X}<0$ and $R$ symmetry is spontaneously broken also.
We refer the reader to \cite{0703196,0902.0030} for details on this subject.

Having a SUSY breaking at hand, 
we introduce inflaton $\phi$ into the model through tri-linear superpotential, 
\begin{eqnarray}{\label{deform}}
W=\kappa_{ij}\phi\phi_{i}\phi_{j}
\end{eqnarray}
due to which the $F$-terms change  as,
\begin{eqnarray}{\label{Fterm2}}
F_{X} &\rightarrow& F_{X}, \nonumber\\
F_{\phi_{i}} &\rightarrow& F_{\phi_{i}}+\kappa_{ij} \phi\phi_{j}, \\
F_{\phi}&\rightarrow& \kappa_{ij}\phi_{i}\phi_{j}. \nonumber
\end{eqnarray}
It is easily to verfy that original SUSY breaking vacuum doesn't be modified for arbitrary $\left<\phi\right>$.
Note that either linear or quadratic interaction in superpotential for inflaton would modify the vacuum in Eq.(\ref{Fterm}).

If the $R$ charge of inflaton is different from $X$,
the SUSY-breaking effect restored in $X$ is forbidden to communicate to inflaton.
Conversely,
if they are the same,
some of messengers in Eq.(\ref{superpotential1}) and Eq.(\ref{deform}) can couple to $X$ and inflaton simultaneously.
Thus, SUSY-breaking effect is communicated to inflaton, 
whose soft mass is the same as structure of Eq.(\ref{scales}) but with $M_{*}=\kappa\left<\phi\right>$ explicitly.
During inflation process,  $\left<\phi\right>$ is super-Plankian as hinted by BICEP2 data, 
which implies that 
\begin{eqnarray}
M_{*}=\kappa\left<\phi\right>\sim \kappa M_{P}.
\end{eqnarray}
If $\kappa_{ij}$ is chosen as $\kappa\sim \mathcal{O}(1)$  
one obtains the correct magnitude of scale $m_{*}$ as desired,
and there is no much fine tuning associated to inflaton mass .

\begin{acknowledgments}
We would like to thank J.-h, Huang for discussion.
The work is supported in part by the Fundamental Research Funds for the Central Universities with Grant No. CQDXWL-2013-015
\end{acknowledgments}

\linespread{1}

\end{document}